\documentclass[aps,amsmath,amssymb,twocolumn,10pt,pra]{revtex4-1}
\usepackage[T1]{fontenc}
\usepackage{graphicx}  
\usepackage[colorlinks=true,linkcolor=blue,citecolor=red,urlcolor=magenta]{hyperref}
\usepackage[dvipsnames]{xcolor}
\usepackage{charter}

\begin{document}  
  
\title{Few strongly interacting fermions of different mass \\ driven in the vicinity of a critical point}
\author{Damian W{\l}odzy\'nski and Tomasz Sowi\'nski} 
\affiliation{Institute of Physics, Polish Academy of Sciences, Aleja Lotnik\'ow 32/46, PL-02668 Warsaw, Poland} 
\date{\today} 
 
\begin{abstract} 
It was recently argued that one-dimensional systems of several strongly interacting fermions of different mass undergo critical transitions between different spatial orderings when the external confinement adiabatically changes its shape. In this work, we explore their dynamical properties when finite-time drivings are considered. By detailed analysis of many-body spectra, we show that the dynamics is typically guided only by the lowest eigenstates and may be well-understood in the language of the generalized Landau-Zener mechanism. In this way, we can capture precisely the dynamical response of the system to the external driving. As consequence, we show that by appropriate tailoring parameters of the driving one can target desired many-body state in a non-infinite time. Our theoretical predictions can be straightforwardly utilized in upcoming state-of-the-art experiments with ultracold atoms.
\end{abstract}  
\maketitle  

\section{Introduction} \label{Sec1}
Critical transitions of quantum many-body systems are one of the most intriguing phenomena which can be explored from many different sides. To their most prominent albeit typical examples belong the quantum phase transitions in which many-body systems rapidly switch between different quantum phases having distinct order parameters \cite{1997SondhiRMP,2010DziarmagaAdvPhys,2011PolkovnikovRMP}. Typically, these transitions are studied in terms of adequate lattice models which aim to capture essential properties of particular solid-state materials \cite{1997SondhiRMP,Sachdev2011QPT}. Underlying theoretical explanation of their universal behavior is delivered in the framework of the renormalization group theory \cite{1974FisherRMP,1975WilsonRMP,1992WhitePRL}. In the context of quantum phase transitions, very often one considers not only adiabatic transitions through the critical point but also finite-time quenches which expose further curious features like the Kibble-Zurek scaling \cite{1976KibbleJPhysA,1985ZurekNature}, relaxation and thermalization of closed systems \cite{1991DeutschPRA,1994SrednickiPRE,2009RigolPRL}, or violation of adiabatic invariants conservation \cite{2009AltlandPRA,2009ItinPRA}. Importantly, these studies are not purely theoretical but have also strong experimental support. Let us mention here two seminal examples from the ultracold atomic physics: {\it (i)} observation of the Mott transition triggered by the varying depth of the optical lattice containing Bose-Einstein-condensed $^{87}$Rb atoms \cite{2002GreinerNature} and {\it (ii)} rapid switch between polar and ferromagnetic phases of bosonic spinor condensates \cite{2006SadlerNature}.
 
Quite recently, due to tremendous progress in experimental control on ultracold atomic systems, increasing interest in the few-body sector is clearly visible. In this matter, a groundbreaking milestone was delivered by Jochim's group by obtaining a well-controlled two-component, effectively one-dimensional, a few-fermionic mixture of $^6$Li atoms \cite{2011SerwaneScience,2012ZurnPRL,2013WenzScience}. These experiments have been followed by extensive theoretical analysis in many different directions (see \cite{2012BlumeRPP,2016ZinnerEPJ,2019SowinskiRPP} for reviews). In the light of current experimental possibilities on preparing fermionic macroscopic mixtures of different elements \cite{2008WillePRL,2010TieckePRL,2016CetinaScience,2018RavensbergenPRA} few-body mixtures of different mass atoms are also theoretically considered as attainable in the near future \cite{2013CuiPRL,2015LoftEPJD,2016PecakNJP,2017HarshmanPRX,2019MistakidisNJP}, including also dynamical problems when the system is out of a stationary state or it is driven by a time-dependent disturbances \cite{2016EberAnnPhys,2018ErdmannPRA,2019ErdmannPRA,2019BarfknechtSciRep,2020NandyNJP}. 

Typically, one associates critical transitions with many-body systems being close to their thermodynamic limit, {\it i.e.}, when the number of particles and spatial sizes of the system are extended to infinity (keeping their ratio constant). In this limit, when external control parameters approach the critical point, one observes that some of the system's features display power-law divergent with well-defined critical exponents. It is said that the considered system becomes critical. However, it should be pointed out that granting the thermodynamic limit is not in fact a necessary condition to make a system critical. One another possibility was found recently for two-component one-dimensional systems of a few strongly interacting fermions \cite{2016PecakPRA,2020WlodzynskiPRA}. It was argued, that in the vicinity of infinite inter-component repulsions, along with changing shape of external confinement, the many-body ground state undergoes a critical transition between two different spatial orderings. Importantly, this transition is characterized by divergent behavior of different observables having appropriate scaling properties when interactions, instead of the size of the system, tend to infinity. The analogy to typical quantum phase transitions was supported further in \cite{2020WlodzynskiPRA} by confirming that quantum correlations between different parties of the system, quantified by appropriate entanglement entropies, are also power-law divergent in the transition point \cite{2002OsterlohNature,2003VidalPRL,2004QuPRL}.

In previous studies, the critical transition of few-fermion systems was analyzed only from the point of view of their ground-state properties. This approach corresponds to the adiabatic transition for which the shape of the external trap is tuned infinitesimally slow. Since the class of systems considered can be prepared, controlled, and measured in state-of-the-art experiments \cite{2008WillePRL,2011SerwaneScience,2012ZurnPRL,2013WenzScience,2010TieckePRL,2016OnofrioUFN} it is important to know their properties not only in this theoretical, experimentally unattainable limit but also when finite-time transitions are examined. In this work, we consider the simplest scenario of a single period of driving of the system containing four particles (balanced and imbalanced) initially prepared in its interacting many-body ground state and being close to the critical point. We focus on sub-critical systems, {\it i.e.}, interactions are very strong but not infinite. By detailed inspection of the many-body spectrum, we show that for finite (but slow enough) drivings properties of the system are well-captured by taking into account only the lowest many-body eigenstates. Particularly simple are systems of an imbalanced number of particles -- their dynamical properties are determined by the time evolution of only two many-body eigenstates. Thus, these systems can be described by a straightforward generalization of the two-level Landau-Zener model \cite{1932Landau,1932Zener}. In this way systems containing a mesoscopic number of particles form a very interesting connection between the simplest quantum two-level systems and very complex strongly interacting many-body systems undergoing critical transitions.

Our work is organized as follows. In Sec.~\ref{Sec2} we describe the system studied, discuss its basic properties, and define the driving protocol used in further analysis. Then in Sec.~\ref{Sec3} we explore temporal many-body spectra of balanced and imbalanced systems and we describe their properties in the adiabatic limit, {\it i.e.}, when driving is infinitely slow and the system remains in its temporal ground state. In Sec.~\ref{Sec4}, by defining transition probabilities, we expose dynamical consequences of finite-time driving. We show that depending on the balance between components, the system can be well-approximated by a two- or four-level model. In the former case, the results are in perfect agreement with those predicted by the approximate method originating in the Landau-Zener model. For completeness of the analysis, in Sec.~\ref{Sec5} we discuss the behavior of the system from the quantum correlations point of view. We show that with appropriate tailoring of driving parameters one can control amount of quantum correlations gained during the evolution. Finally, we summarise in Sec.~\ref{Sec6}.

\section{The system studied} \label{Sec2} 
In this paper, we consider an ultracold two-component, mass-imbalanced mixture of a few fermions confined in a tight one-dimensional trap. Interactions between fermions are assumed to be dominated by the $s$-wave scattering and modeled by the $\delta$-like potential. The Hamiltonian of the system in the second quantization formalism has a form:
\begin{align} \label{Hamiltonian}
\hat{\cal H} &= \sum_{\sigma\in\{A,B\}}\int\!\mathrm{d}x\,\hat{\Psi}^\dagger_\sigma(x)\left[-\frac{\hbar^2}{2m_\sigma}\frac{\mathrm{d}^2}{\mathrm{d}x^2}+V_\sigma(x)\right]\hat\Psi_\sigma(x) \nonumber \\
&+ g\int\!\mathrm{d}x\,\hat{\Psi}^\dagger_A(x)\hat{\Psi}^\dagger_B(x)\hat{\Psi}_B(x)\hat{\Psi}_A(x),
\end{align}
where indexes $A$ and $B$ denote lighter and heavier component of the mixture, respectively. A fermionic field operator $\hat{\Psi}_\sigma(x)$ annihilate particle from component $\sigma$ at position $x$ and fulfill standard anti-commutation relations, $\{\Psi_\sigma(x),\Psi^\dagger_{\sigma'}(x')\}=\delta_{\sigma\sigma'}\delta(x-x')$ and $\{\Psi_\sigma(x),\Psi_{\sigma'}(x')\}=0$. It is clear that the Hamiltonian $\hat{\cal H}$ does commute with the particle number operators $\hat{N}_\sigma = \int\!\mathrm{d}x\, \hat{\Psi}^\dagger_\sigma(x) \hat{\Psi}_\sigma(x)$. The inter-component interactions in the system are described by the effective interaction strength $g$ and in all the cases considered in this paper are strongly repulsive. From the experimental point of view, this parameter can be controlled by varying intensity of the external confinement in directions perpendicular to the direction of motion \cite{1998OlshaniiPRL} and/or via Feshbach resonances \cite{2008WillePRL,2010TieckePRL}. In the following, the external confining potential $V_\sigma(x)$ is controlled by a shape parameter $\lambda$ and has a form:
\begin{align} \label{potential} 
V_\sigma(x) = \left\{
\begin{array}{cl}
\frac{1}{2}\lambda m_\sigma x^2, & |x|<L, \\
\infty, & |x|\geq L
\end{array}
\right.
\end{align}
Parameter $\lambda$ has a natural interpretation of a square of the effective frequency in the center of the system. Thus, it is clear that for $\lambda=0$ the potential has a form of a uniform box, while for large enough $\lambda$ ({\it i.e.}, $\lambda\gg\hbar^4/m^2_\sigma L^4$) it has a form of a harmonic trap cropped at the edges by a hard walls. As was argued previously in \cite{2016PecakPRA}, whenever repulsive interactions are strong enough, the system is forced to separate its components. A particular separation scenario depends on the shape of an external potential. In the case of box potential ($\lambda=0$), the single-particle density profile of heavier particles is split and pushed out to the edges of the trap while the lighter component gathers near its center. Contrary, in the harmonic potential (large enough $\lambda$), the lighter component is split and located on the edges while the heavier one remains in the center. Consequently, when the parameter $\lambda$ is changed between these two regimes the system undergoes some sort of transition \cite{2016PecakPRA,2020WlodzynskiPRA} whose properties are quite similar to that known from typical quantum phase transitions. Namely, actual quantities characterizing phases on opposite sides of the transition point become divergent (with appropriate scaling) when the system approaches the transition point, $\lambda_0$. Exactly at this point, in the limit of infinite repulsions, the system becomes critical.  

In our work, we want to make the first step beyond the adiabatic transition and examine the properties of the system when the shape of the potential is suitably modulated. To make this analysis as simple as possible we focus on sub-critical (very strong but finite repulsions) systems with $N_A+N_B=4$ particles with a mass ratio $m_B/m_A=40/6$ corresponding to the Li-K atomic mixture. We assume that initially the system is confined in the box potential ($\lambda=0$) and prepared in its many-body interacting ground state, $|\Psi(0)\rangle\equiv|\mathtt{0}\rangle$. Then it is driven forward and backward through the transition point. In our approach we assume that the shape parameter $\lambda$ changes periodically as:
\begin{align} \label{lambda}
\lambda(t) = 2 \lambda_0 \sin^2\left(\frac{\pi t}{\tau}\right),
\end{align}
where $\lambda_0$ is the point of critical transition (its particular value depends on the number of particles and detailed shape of the external potential) and $\tau$ is a time period after which the parameter $\lambda$ returns to its initial value. We suspect that in the limit $\tau\rightarrow\infty$ the state of the system continuously remains in the temporal many-body ground state and thus previous results for the adiabatic transition are restored. Since in practice, all considered scenarios happen far from the limit of a purely harmonic trap, it is convenient to perform analysis in natural units of box potential, {\it i.e.}, units of length, energy, frequency, and interaction strength are $L$, $\hbar^2/(m_A L^2)$, $\hbar/(m_A L^2)$, and $\hbar^2/(m_A L)$, respectively. In this units we set interaction strength $g = 20$.

\section{Adiabatic limit} \label{Sec3}
\begin{figure}
\centering
\includegraphics[width=\linewidth]{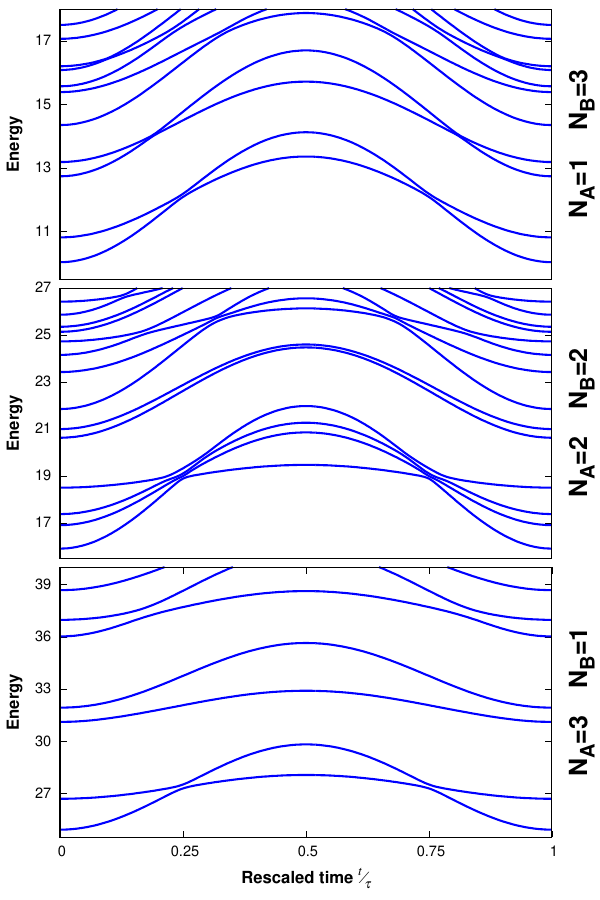}
\caption{Temporal spectra of the Hamiltonian \eqref{Hamiltonian} in the subspace of the parity operator $\hat{\boldsymbol{\mathsf P}}$ containing the initial state of the system $|\Psi(0)\rangle$ for different number of particles. Note a characteristic quasi-degeneracy of the lowest states which appears when the system approaches vicinity of the transition point ($t/\tau \approx 0.25$ and $0.75$). Depending on the balance of particles, the ground manifold two or four eigenstates.\label{Fig1}}
\end{figure}

Before we discuss the quench dynamics of the system let us first describe properties of the many-body spectrum which reflects the system's properties during the adiabatic dynamics ($\tau\rightarrow \infty$). It is a matter of fact, that all the eigenstates of the Hamiltonian (\ref{Hamiltonian}) can be divided into two disjoint classes. These two classes correspond to two orthogonal subspaces of the spatial parity operator $\hat{\boldsymbol{\mathsf P}}:x\rightarrow -x$. Indeed, since the external potential remains always spatially-symmetric (one finds $V_\sigma(x)=V_\sigma(-x)$ for any $\lambda$) and the contact inter-component interactions do not change corresponding parity of the many-body Fock states, one can show that the many-body Hamiltonian $\hat{\cal H}$ commutes with $\hat{\boldsymbol{\mathsf P}}$. In consequence, the Hamiltonian can be diagonalized in each eigensubspace of $\hat{\boldsymbol{\mathsf P}}$ independently. Moreover, since initially the system is prepared in the many-body ground state belonging to one of these subspaces, successive evolution of the system (independently on the details of the protocol $\lambda(t)$) is restricted to this eigensubspace. Therefore, in the following we always restrict ourselves to the eigensubspace of $\hat{\boldsymbol{\mathsf P}}$ which contains the initial many-body state, {\it i.e.}, $|\Psi(0)\rangle=|\mathtt{0}\rangle$. Technically, the eigenspectrum of the Hamiltonian is calculated by a direct numerical diagonalization. The Hamiltonian is represented as a matrix in the Fock basis build from $K=32$ the lowest single-particle orbitals of the noninteracting system with $\lambda=0$. We checked that the results obtained with this cut-off are well-converged, {\it i.e.}, they do not change significantly when higher cut-offs are employed. In Fig.~\ref{Fig1} we display spectra (in the chosen eigensubspace) of the Hamiltonian $\hat{\cal H}$ for different distributions of particles as functions of time rescaled by the transition period $\tau$. It is clear that for any shape $\lambda$, each many-body eigenstate of the system remains isolated from other states. It automatically means that in the case of the adiabatic driving the system remains in the temporal ground state of the Hamiltonian and finally returns to its initial form. We note, however, that for confinements being close to the transition ($t/\tau \approx 0.25$ and $0.75$), energy gaps between particular eigenstates decrease, and corresponding states bunch to quasi-degenerate manifolds. At the same time, energy gaps between different manifolds persist largely (for these particular systems the value of the shape parameter $\lambda_0$ is $0.55$, $0.81$, and $1.07$, respectively for systems with $N_A=4-N_B=1,2$, $3$ particles). It means that for slow enough transitions, the evolution of the system is guided only by a first few eigenstates. Only in the limit of infinite repulsions (do not consider here) these energy gaps vanish and spectra become degenerated. At this point let us also emphasize a fundamental difference between the balanced system of $N_A=N_B=2$ particles and unbalanced ones with $|N_A-N_B|=2$. As shown in Fig.~\ref{Fig1}, when the system is close to the transition, in the former case four eigenstates become quasi-degenerated in the ground manifold, while in the latter only two of them approach degeneracy. Of course, this substantial qualitative difference between balanced and imbalanced systems will have an impact on the dynamical properties of the system subjected to quench dynamics. As a side note let us also mention that typically the degeneracy of manifolds appearing in the limit of infinite interactions is essentially different than the degeneracy for systems containing particles of the same mass. For equal mass systems degeneracy of the ground manifold (in the $g\rightarrow\infty$ limit) is independent of the external confinement and it is always equal $(N_A+N_B)!/(N_A!\,N_B!)$ ~\cite{2009GuanPRL}. Contrary, when particles belonging to different components have different masses, this degeneracy is partially lifted and becomes dependent on a shape of external confinement. 

In the case of non-adiabatic transitions described below, an exact evolution of the system is predicted by solving the many-body Schr\"odinger equation $i\hbar \partial_t |\Psi(t)\rangle = {\cal H}(t)|\Psi(t)\rangle$ straightforwardly. We represent the Hamiltonian as a time-dependent matrix in the time-independent Fock basis described above. Then we use the fourth-order Runge-Kutta method with time step $\delta t$ at least $10^{-2}$ times smaller than $\tau$ and at least $10^{-2}$ times smaller than $\hbar/\delta E$, where $\delta E$ is the largest relevant gap in the spectrum. We checked that this accuracy is sufficient for obtaining well-converged results.

\section{Transition probability} \label{Sec4}

One of the simplest quantities characterizing properties of the driven system is the temporal probability ${\cal P}_0(t)=|\langle\mathtt{0}(\lambda(t))|\Psi(t)\rangle|^2$ that the system remains in its current ground state $|\mathtt{0}(\lambda(t))\rangle$. Of course, we suspect that in the limit of very long transition periods ($\tau\rightarrow\infty$), {\it i.e.}, when the transition resembles the adiabatic process, the state follows its temporal ground state. Thus, in this limit the probability saturates at unity, ${\cal P}_0(t)\rightarrow 1$. It is clear that for faster transitions the probability ${\cal P}_0$ is typically less than one and some other probabilities ${\cal P}_n(t)=|\langle\mathtt{n}(\lambda(t))|\Psi(t)\rangle|^2$ increase signalling that the system can be detected in $n$-th temporal eigenstate $|\mathtt{n}(\lambda(t))\rangle$. 

From both, experimental and theoretical, points of view, it is fundamentally important to settle if one can prepare a finite-time transition in such a way that at the final instant the selected probability becomes exactly equal to one. Specifically, the famous {\it shortcut to adiabaticity problem} \cite{2019OdelinRMP,2021ChenPRL} is related to finalizing the dynamics of the system in its many-body ground state, ${\cal P}_0(\tau)=1$. To find a satisfying answer to this question, let us focus on two well-defined instants of the transition process: the moment just after the first transition, $t= \tau/2$, and the moment after a whole cycle of driving when the Hamiltonian returns to its initial form, $t=\tau$. 

\begin{figure}
\centering
\includegraphics[width=\linewidth]{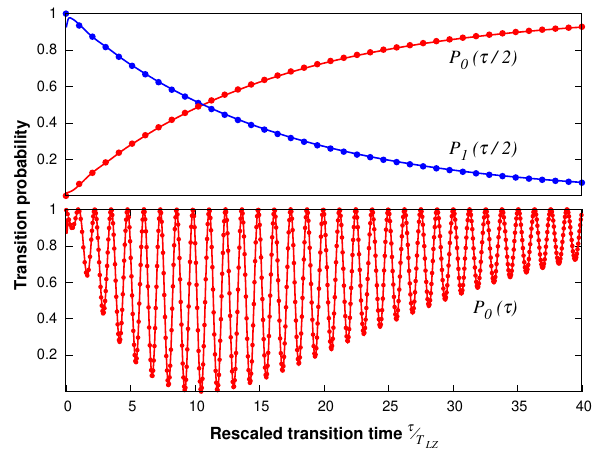}
\caption{Transition probabilities ${\cal P}_0(\tau/2)$ and ${\cal P}_1(\tau/2)$ for the system of $N_A=1$ and $N_B=3$ particles after the first transition {\bf (top panel)} and the transition probability ${\cal P}_0(\tau)$ after a whole driving period {\bf (bottom panel)} as functions of the transition time $\tau$. Red and blue solid lines correspond to the numerically exact results obtained by solving the many-body Schr\"odinger equation straightforwardly. Corresponding dots are obtained in the Landau-Zener approximation without any fitting parameters. Note a small deviations from the Landau-Zener approximation for sufficiently rapid transitions ($\tau\rightarrow 0$). \label{Fig2}}
\end{figure}

First, let us discuss the simpler cases of the imbalanced system of $|N_A-N_B|=2$ particles. Since both imbalanced cases have very similar structures of the many-body spectrum, their dynamical properties caused by the transition protocol $\lambda(t)$ are qualitatively the same. Therefore, in the following, we focus on the particular case of $N_A=1$ and $N_B=3$ particles. The half-time and the final transition probabilities ${\cal P}_n(\tau/2)$ and ${\cal P}_n(\tau)$ as functions of the transition time $\tau$ are shown in Fig.~\ref{Fig2}. It is clear that after the first transition ($t=\tau/2$) the system may become excited to another many-body state $|\mathtt{1}\rangle$. Moreover, if the transition is instantaneous ($\tau\rightarrow 0$) then the system is almost perfectly transformed to this state, ${\cal P}_0(\tau/2)\approx 0$. In contrast, in the adiabatic limit ($\tau\rightarrow \infty$) the system continuously remains in a temporal many-body ground state, and ${\cal P}_0(\tau/2)$ remains equal to 1. For intermediate transition periods $\tau$ both probabilities smoothly interpolate between these two scenarios. This overall picture is in perfect agreement with predictions of the Landau-Zener model describing a two-level system undergoing finite-time transition \cite{1932Landau,1932Zener}. Indeed, taking into account dependence of the energy spectrum displayed in Fig.~\ref{Fig1} one can straightforwardly show that the transition probability to the excited state after the first transition is expressed as 
\begin{equation}
{\cal P}_1(\tau/2)=1-{\cal P}_0(\tau/2)=\mathrm{exp}\left(-\pi \sqrt{\frac{2\delta}{\gamma}} \frac{\tau}{T_{LZ}} \right),
\end{equation}
where $\delta$ and $\gamma$ are directly related to properties of the spectrum at the transition point $\lambda=\lambda_0$. Namely, if the energy difference between two eigenstates is written as $\Delta{\cal E}(t)=E_1(t)-E_0(t)$ then $2\delta=\Delta{\cal E}(t)|_{t=\tau/4}$ and $\gamma=\partial^2_t\Delta{\cal E}(t)|_{t=\tau/4}$. The half-energy gap $\delta$ defines a natural Landau-Zener time scale $T_{LZ}=\hbar/\delta$. To make this quantity more understandable, let us relate it to possible experimental realisations. For example, if we consider a typical experiment with BEC confined in a potential box of width $L\sim 80\,\mathrm{\mu m}$ \cite{2005MeyrathPRA,2013GauntPRL}, the corresponding Landau-Zener time $T_{LZ}$ is quite large and of the order of several seconds ({\it i.e.}, $12.1~\mathrm{s}$, $11.1~\mathrm{s}$, and $5.4~\mathrm{s}$ for $N_A=1,2$ and $3$, respectively). However, if we consider the Heidelberg experiments where spatial sizes of confining traps are around 20 times smaller, {\it i.e.}, around one micrometer per particle, then the Landau-Zener time is reduced to several milliseconds ({\it i.e.}, $30.3~\mathrm{ms}$, $27.7~\mathrm{ms}$, and $13.4~\mathrm{ms}$ for $N_A=1,2$ and $3$, respectively) making the protocol proposed attainable. Moreover, since $T_{LZ}$ depends directly on the size of the box, it can be experimentally tuned by modifying length $L$.

Taking values of $\delta$ and $\gamma$ directly from the energy spectrum obtained numerically, we see that our numerically exact predictions are in perfect agreement with those obtained in the framework of the Landau-Zener theory (solid lines and dots in Fig.~\ref{Fig2}, respectively). Note however that there is a clear deviation from the Landau-Zener approximation for very rapid transitions ($\tau\rightarrow 0$). This is a direct consequence of the fact that two the lowest many-body states at $t=0$ do not span exactly the same space as states at $t=\tau/2$ but rather they have also tiny contributions from higher many-body states. Thus, for rapid transition the state $|\mathtt{0}(\lambda(0))\rangle$ is not transit ideally to the state $|\mathtt{1}(\lambda(\tau/2))\rangle$ and the transition ${\cal P}_1(\tau/2)$ becomes lower than 1 although ${\cal P}_1(0)$ remains equal 0.

The situation changes slightly when the system is driven back again through the second, returning transition. As clearly seen (bottom panel in Fig.~\ref{Fig2}), in this case, the probabilities ${\cal P}_0(\tau)$ and ${\cal P}_1(\tau)$ experience specific oscillations. Importantly, by an appropriate choice of the {\it finite} transition time $\tau$ it is possible to drive back the system ideally to the ground state. Explanation of this non-monotonic behavior is not complicated but requires full quantum-mechanical argumentation. In general, after the first transition, the system is in some well-defined superposition of the two many-body states. Thus, the probability that the system will return to the ground state after the second transition cannot be expressed as simple products of half-time probabilities ${\cal P}_0(\tau/2)$ and ${\cal P}_1(\tau/2)$ (like the naive semi-classical picture suggests) but it requires additional knowledge of the relative phase $\phi_{\tau/2}$ of the half-time superposition of many-body eigenstates. Careful but straightforward considerations of this issue, similarly as it is done in the theory of Landau-Zener-Stueckelberg interferometry \cite{2010ShevchenkoPhysRep}, shows that the final probability of returning to the ground state is expressed as
\begin{equation} \label{Pphi}
{\cal P}_0(\tau) = \left|{\cal P}_0(\tau/2)\mathrm{e}^{2i\phi_{\tau/2}} + {\cal P}_1(\tau/2)\right|^2.
\end{equation}
Since ${\cal P}_0(t)+{\cal P}_1(t)=1$, the final probability ${\cal P}_0(\tau)$ is equal to one whenever the relative half-time phase $\phi_{\tau/2}$ is multiple of $\pi$. Obviously, the phase $\phi_{\tau/2}$ depends on the details of the many-body spectrum during a whole evolution, the detailed form of the transition protocol $\lambda(t)$, and the transition time $\tau$. Nevertheless, it can be determined rigorously if these features are known. Namely, as shown in \cite{Kayanuma_1994}, the phase $\phi_{\tau/2}$ depends on the transition time $\tau$ as $\phi_{\tau/2}=A(\tau) + B\tau$, where $B$ is transition-time-independent constant determined by the spectrum, while $A(\tau)$ is a monotonic function interpolating between two extreme cases $A(\tau=0)=\pi/4$ and $A(\tau\rightarrow\infty)=0$. The exact form of this relation is given in \cite{Kayanuma_1994}. Taking all these considerations into account, we find that the relation \eqref{Pphi} ideally reproduces our numerical results (solid lines and dots in Fig.~\ref{Fig2}) without any fitting parameters. 

\begin{figure}
\centering
\includegraphics[width=\linewidth]{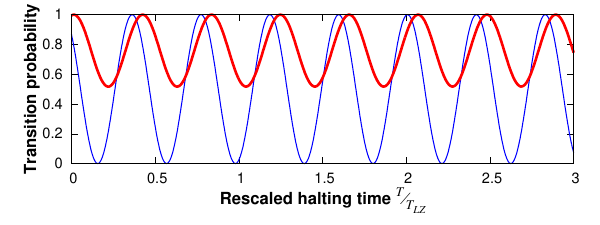}
\caption{The final transition probability ${\cal P}_0$ for the system of $N_A=1$ and $N_B=3$ particles as a function of the halting time $T$ for two representative driving periods $\tau = 10 T_{LZ}$ (thin blue) and $\tau = 30 T_{LZ}$ (thick red). It is clear that by appropriate choice of the halting time $T$ one can easily tune the final transition probability and thus the final state of the system. Importantly, it is not always possible to tune the halting time to ensure a transition to the excited state. See the main text for details.\label{Fig3}}
\end{figure}

It is worth noticing that the half-time phase $\phi_{\tau/2}$ can be controlled quite easily by a simple interruption of the protocol at $t=\tau/2$ for some particular time $T$. Due to the substantial difference of eigenenergies at this moment, the relative phase is ballistically winded by a factor $\delta\phi = \Delta{\cal E}(\tau/2) T/\hbar$. Thus, it can be tuned to any chosen value by the appropriate choice of the halting time $T$. In Fig.~\ref{Fig3} we display the dependence of the final probability ${\cal P}_0(\tau)$ on the halting time $T$ for two exemplary transition times $\tau$. For unfrozen protocol ($T=0$) the final probability ${\cal P}_0$ is the same as determined previously. It is clear that by changing the halting time $T$, the final probability ${\cal P}_0$ can be easily tuned to unity. Note however that it is not always possible to tune the probability to $0$, {\it i.e.}, to assure a perfect transition to the excited state. This impossibility is caused by specific values of half-time probabilities ${\cal P}_0(\tau/2)$ and ${\cal P}_1(\tau/2)$ which are determined solely by $\tau$. It is worth mentioning that exactly this idea was recently exploited in quantum dot systems to tailor a relative phase and in consequence, the final state of the system to the desired form \cite{2010PettaScience}. Thus, the method presented can be viewed as an approach for quantum optimal control and shortcut to adiabaticity techniques complementary to a variety of other methods based on non-adiabatic change of the control parameters \cite{2010SchaffPRA,2012BasonNatPhys,2015RohringerSciRep,2016DuNatureComm,2016FrankSciRep,2018DengPRA}.

\begin{figure}[h!]
\centering
\includegraphics[width=\linewidth]{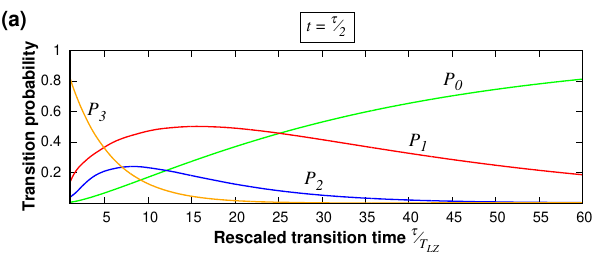}\\
\includegraphics[width=\linewidth]{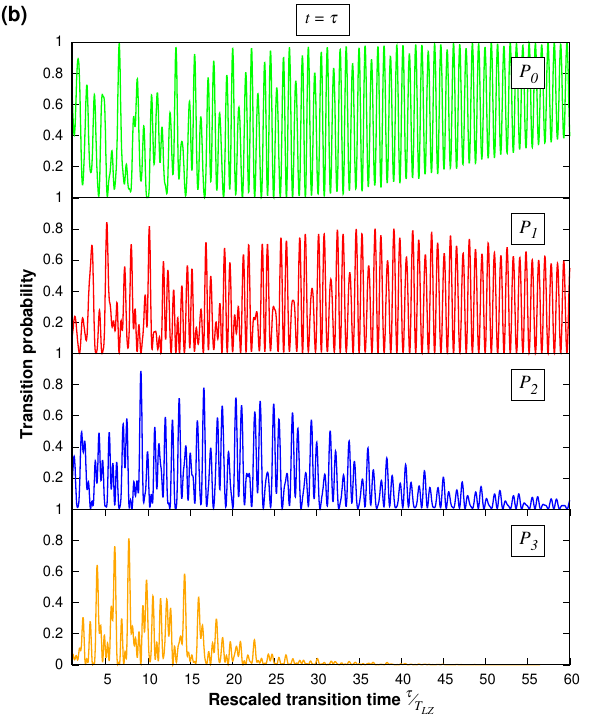}\\
\includegraphics[width=\linewidth]{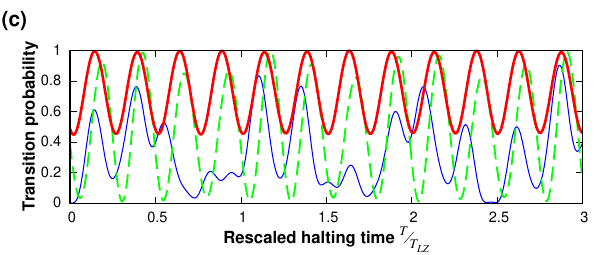}
\caption{Transition probabilities ${\cal P}_n(t)$ for the balanced system of $N_A=N_B=2$ particles {\bf (a)} after the first transition ($t=\tau/2$), {\bf (b)} after a whole driving period ($t=\tau$) as functions of the transition time $\tau$. Different colors correspond to different transition probabilities ${\cal P}_n(t)$. {\bf (c)} Transition probability ${\cal P}_0$ of returning to the ground state after the whole evolution as a function of the halting time $T$ for three different driving periods $\tau/T_{LZ}=10$ (thin blue), $\tau/T_{LZ}=30$ (dashed green), $\tau/T_{LZ}=60$ (thick red).\label{Fig4}}
\end{figure}

Of course, the dynamical properties of the balanced system of $N_A=N_B=2$ particles are substantially different. It is a direct consequence of the structure of the system's eigenspectrum having not two but four different many-body states (we enumerate them with $n\in\left\{0,1,2,3\right\}$, respectively). Depending on the driving time $\tau$, different states are occupied with different probabilities. Nonetheless, the overall picture just after the first transition ($t=\tau/2$) is still relatively simple. As shown in Fig.~\ref{Fig4}a, occupations of four relevant states at half-time instant ${\cal P}_n(\tau/2)$ are sensitive to the transition time. Note that for sufficiently large $\tau$ only two of the lowest states contribute to the dynamics. Thus, in this case, the system resembles the imbalanced scenario. This behavior is in full agreement with phenomenological explanation -- for sufficiently slow processes the system follows its temporal ground state and the most relevant corrections are induced mostly by excitations to the nearest many-body state. Therefore, a two-level approximation is sufficient to explain the system's behavior.

After a whole driving cycle ($t=\tau$), occupations ${\cal P}_n(\tau)$ of different many-body states become much more complicated (Fig.~\ref{Fig4}b). Detailed but straightforward analysis shows that analogously to the imbalanced case they are dependent on their corresponding half-time values ${\cal P}_n(\tau/2)$ and {\it three} different phases related to the half-time superposition of the system's state. For example, the probability of returning the system to its initial ground state after a whole period is given by (compare with \eqref{Pphi})
\begin{equation}
{\cal P}_0(\tau) = \left|{\cal P}_0(\tau/2)+\sum_{n=1}^3{\cal P}_n(\tau/2)\mathrm{e}^{2i\phi_n}\right|^2.
\end{equation}
These three different and independent phases are the main sources of irregular oscillations of the final probabilities when plotted as functions of the transition time $\tau$. However, for sufficiently slow driving (large enough $\tau$) only two of the lowest states are occupied. As mentioned earlier, in this range of $\tau$ reduction of the problem to the two-level system is fully relevant and the description becomes analogous to the imbalanced counterpart. When the transition time $\tau$ is increased further, the contribution of the lowest excited state slowly diminishes. Finally, in the adiabatic limit ($\tau\rightarrow 0$), the system remains in its initial ground state. For rapid transitions, all four states contribute to the final state of the system. Although transition probabilities are very irregular, still it is possible to choose transition time $\tau$ to perform a perfect return to the initial ground state. Importantly, this tuning is possible not only in the range of slow transitions where two many-body states contribute to the dynamics but also for quick quenches where the non-trivial interplay of all four eigenstates is less trivial.

For completeness of the discussion, let us mention that also in the balanced-system scenario one has a possibility to tune final probabilities by a transient halting of the driving at $t=\tau/2$. In this case, however, this control has limited capability due to non-trivial relations between transition time $\tau$, the relative phases $\phi_n$, and the half-time probabilities ${\cal P}_n(\tau/2)$ (see Fig.~\ref{Fig4}c for particular examples).

\section{Control of quantum correlations} \label{Sec5}

Dynamical control of the target quantum state of the system described above is only one of interesting directions of quantum engineering. Equally important is an ability to control quantum correlations gained by the system during the evolution. Since the system considered is strongly interacting, it is interesting to ask the question if one can drive the system in such a way that the final state $|\Psi(\tau)\rangle$ is the most or the less correlated \cite{2021ChenPRL}. Since here we deal with the two-component mixture of distinguishable components it is natural to focus on the inter-component correlations which can be in principle determined experimentally \cite{2005WiesniakNJP,2012AbaninPRL,2015IslamNature,2016HaukeNatPhys}. These correlations are typically quantified with the von Neumann entanglement entropy defined as
\begin{equation}
{\cal S}(t) = -\mathrm{Tr}\left[\hat\rho_A(t)\,\mathrm{ln}\hat\rho_A(t)\right]=-\mathrm{Tr}\left[\hat\rho_B(t)\,\mathrm{ln}\hat\rho_B(t)\right],
\end{equation}
where $\hat\rho_\sigma(t)$ is the reduced density matrix of $\sigma$ component obtained from a temporal state of the whole system $|\Psi(t)\rangle$ by tracing out the remaining component $\sigma'$, $\hat\rho_\sigma(t) = \mathrm{Tr}_{\sigma'}\left(|\Psi(t)\rangle\langle\Psi(t)|\right)$. In the following we focus on properties of entanglement entropy after a whole period of the driving, ${\cal S}(\tau)$.  

The case of adiabatic transition ($\tau\rightarrow\infty$) can be easily reconstructed from detailed studies on the ground-state properties of the system \cite{2020WlodzynskiPRA}. It is known that in this case, the entropy rapidly increases near the critical shape of the trap, signaling a strong enhancement of inter-component correlations. Then it decreases and finally saturates on a value appropriate for the ground state on the opposite side of the transition point. Of course, after the reverse process, as suspected, the entropy comes back to its initial value. This picture changes significantly when the transition is not adiabatic since then other many-body states start to contribute. It is clear that quantum inter-component correlations depend on equal footing on the final superposition and the internal structure of each participating many-body state. Therefore, even in the imbalanced case of four particles, {\it i.e.}, when only two many-body states are relevant, the entropy may not be trivial.

\begin{figure}
\centering
\includegraphics[width=\linewidth]{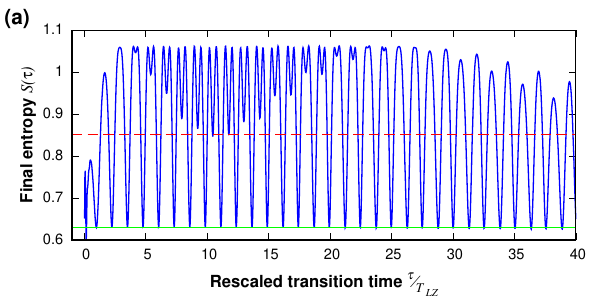}
\includegraphics[width=\linewidth]{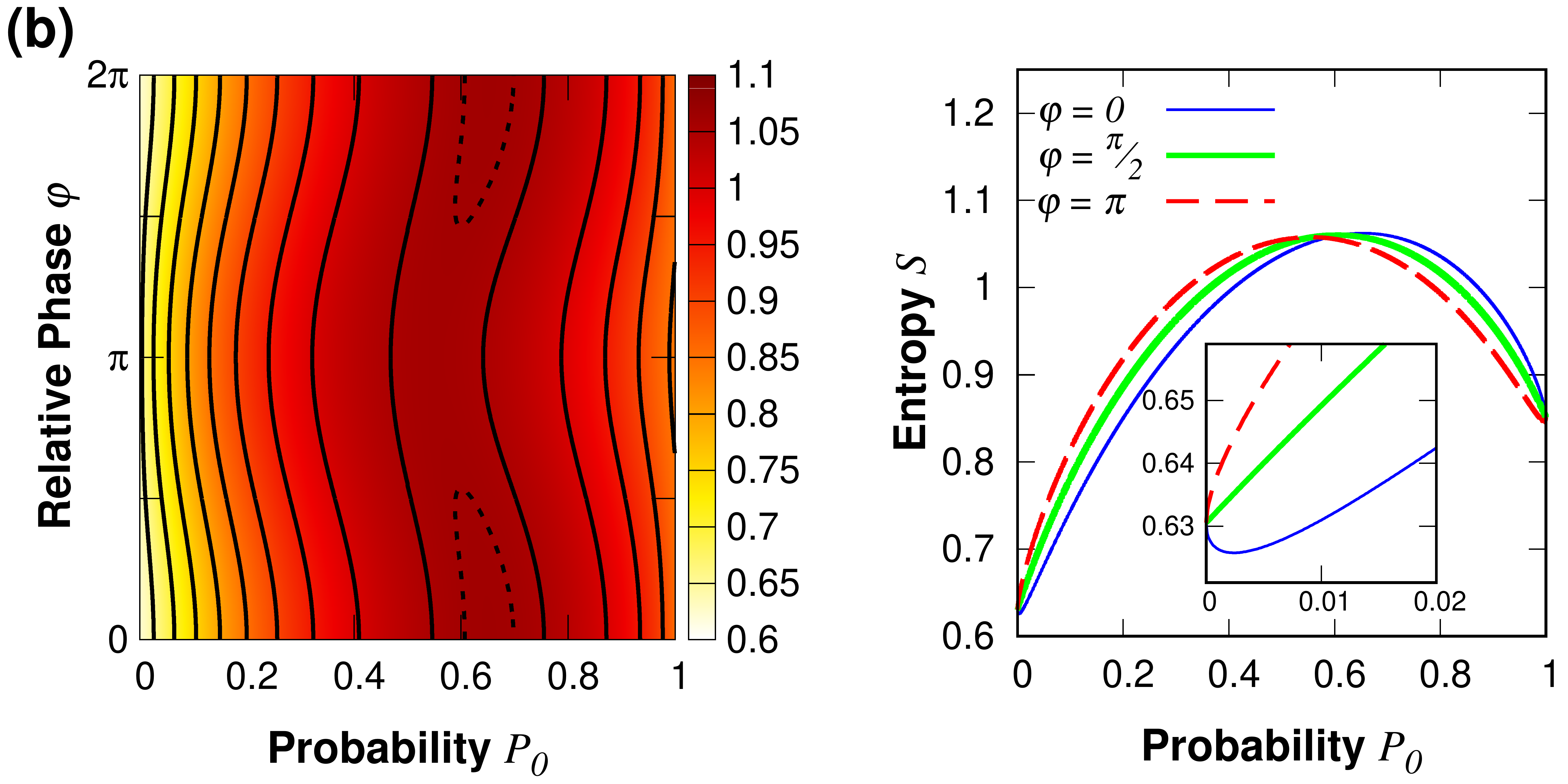}
\caption{{\bf (a)} Inter-component entanglement entropy ${\cal S}(\tau)$ at final instant as a function of the driving period $\tau$ for a system of $N_A=1$ and $N_B=3$ particles. Horizontal solid green and dashed red lines indicate values of the entanglement entropy ${\cal S}$ in many-body eigenstates $|\mathtt{0}\rangle$ and $|\mathtt{1}\rangle$, respectively. {\bf (b)} Entanglement entropy as a function of superposition parameters ${\cal P}_0$ and $\varphi$ between two eigenstates of the Hamiltonian of the same system. The minimal entropy is obtained for the system being close to the ground state $|\mathtt{0}\rangle$, while the maximally entanglement state is a superposition described by parameters ${\cal P}_0\approx 0.65$ and $\varphi=0$. \label{Fig5}}
\end{figure}

Let us start from the system of $N_A=1$ and $N_B=3$ particles. Based on previous considerations, it is clear that the system at the final moment is in the superposition of two many-body states with amplitudes determined by occupations ${\cal P}_0(\tau)$ and ${\cal P}_1(\tau)$. One can write this superposition as
\begin{equation}
|\Psi(\tau)\rangle = \sqrt{{\cal P}_0(\tau)}|\mathtt{0}\rangle + \sqrt{{\cal P}_1(\tau)}\mathrm{e}^{i\varphi(\tau)}|\mathtt{1}\rangle,
\end{equation}
where $\varphi(\tau)$ is the relative phase between the states after the full period of driving. The resulting entropy as a function of driving time $\tau$ is displayed in Fig.~\ref{Fig5}. Although the general dependence of ${\cal S}(\tau)$ resembles the behavior of the final probability ${\cal P}_0(\tau)$ (compare with Fig.~\ref{Fig2}), some substantial differences are also visible. The main difference comes from the fact that extreme values of the entropy are not achieved for any of the system's eigenstates (indicated in Fig.~\ref{Fig5} by horizontal green and red solid lines) but rather for their particular, well-defined, superpositions. In consequence, the largest (lowest) quantum correlations are not obtained for the same driving periods $\tau$ as the largest (lowest) transition probabilities ${\cal P}_n(\tau)$. Especially this is evidently the case when the maximally entangled state is considered. This is reflected in Fig.~\ref{Fig5} as an additional periodic structure close to the bound of maximal entropy. It should be pointed, however, that the many-body eigenstate $|\mathtt{0}\rangle$ is very close to the state having the lowest inter-component correlations. Therefore, whenever the final probability ${\cal P}_0(\tau)$ is close to one, the system achieves almost the smallest entanglement entropy.

To get a better understanding of these results let us analyze more deeply the relation between inter-component entanglement entropy and the superposition parameters ${\cal P}_0(\tau)$ and $\varphi(\tau)$. As clearly seen in Fig.~\ref{Fig5}b, the minimal amount of correlations is achieved for the superposition which is very close to the many-body ground state $|\mathtt{0}\rangle$ having only a small admixture of the excited state $|\mathtt{1}\rangle$. On the other hand, the maximally correlated many-body state is formed by a particular superposition with significant contribution of both eigenstates $|\mathtt{0}\rangle$ and $|\mathtt{1}\rangle$ (${\cal P}_0=1-{\cal P}_1\approx 0.65$ and $\varphi=0$). Note, however, that there are a plethora of other superpositions encoding almost the same amount of inter-component entanglement as the maximally entangled state. Each of these states can be equally easily targeted with appropriately tailored driving. Indeed, by changing $\tau$ one can tune the superposition to the desired ratio. Again, the relative phase $\varphi(\tau)$ (due to a substantial difference between eigenenergies at final instant $\Delta{\cal E}(\tau)$) can be engineered by a simple prolongation of ballistic dynamics after the driving. All it means, that targeting the state with larger inter-component correlations is much easier and seems to be rather a generic scenario. 

\begin{figure}
\centering
\includegraphics[width=\linewidth]{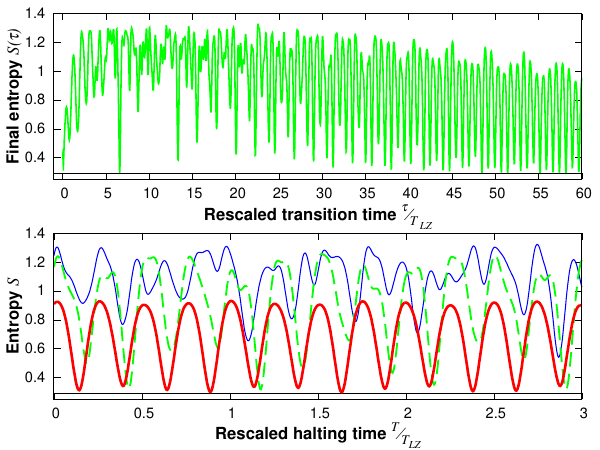}
\caption{{\bf (top panel)} Inter-component entanglement entropy ${\cal S}(\tau)$ at final instant as a function of the transition time $\tau$ for a balanced system of $N_A=N_B=2$ particles. {\bf (bottom panel)} Final entanglement entropy for the same system as a function of the halting time $T$ introduced after the first transition $t=\tau/2$ for three different driving periods $\tau/T_{LZ}=10$ (thin blue), $\tau/T_{LZ}=30$ (dashed green), $\tau/T_{LZ}=60$ (thick red). In both plots, horizontal black solid line indicates the entanglement entropy in the state $|\mathtt{0}\rangle$.\label{Fig6}}
\end{figure}

In the case of the balanced system of four particles, the situation is slightly different. Now, four different many-body states participate in the state at the final instant. Thus, the entropy is strongly affected by different relative occupations ${\cal P}_n(\tau)$ and relative phases $\varphi_n(\tau)$, {\it i.e.}, it is much less regular when plotted as function of $\tau$ (Fig.~\ref{Fig6}). As a consequence, it is much harder (if possible) to tune driving parameters precisely to values maximizing entanglement entropy. In this context, minimizing entropy, by a sufficient prolongation of the transition time $\tau$, seems to be relatively easier since the many-body ground state remains close to the state of minimal inter-component correlations. For completeness, in Fig.~\ref{Fig6}, we also show how the final entanglement entropy ${\cal S}$ depends on the halting time $T$ after the first transition at $t=\tau/2$. It suggests that controlled freezing of the dynamics at half of the period may be an appropriate tool to engineer the final amount of correlations.

\section{Role of the external confinement}
\begin{figure}
\centering
\includegraphics[width=\linewidth]{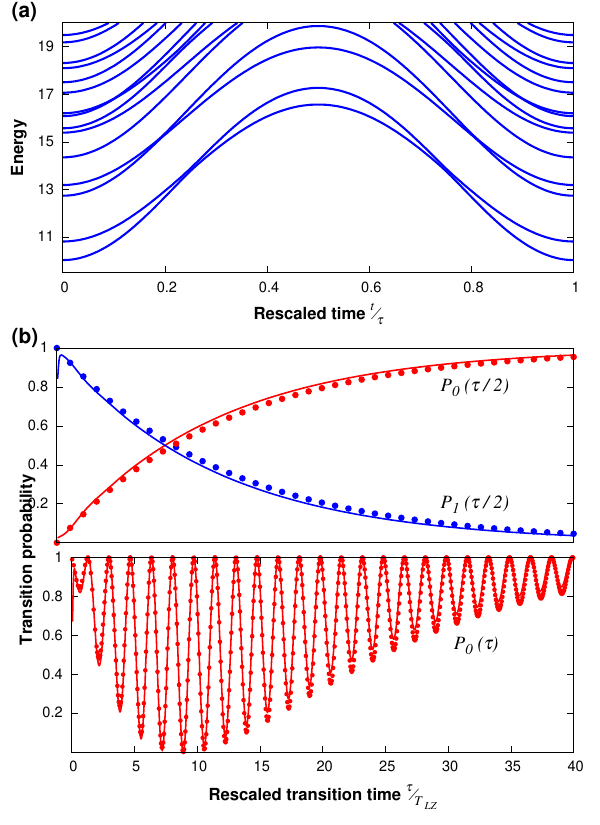}
\caption{{\bf (a)} Temporal spectrum of the system with external potential given by \eqref{potential2} (in the relevant subspace of $\hat{\boldsymbol{\mathsf P}}$) for the system of $N_A=1$ and $N_B=3$ particles. Notice, clear resemblance to the top spectrum presented in Fig.~\ref{Fig1}. {\bf (b)} Transition probabilities ${\cal P}_0(\tau/2)$ and ${\cal P}_1(\tau/2)$ for the same system after the first transition (top) and the transition probability ${\cal P}_0(\tau)$ after a whole driving period (bottom) as functions of the transition time $\tau$. Red and blue solid lines correspond to the numerically exact results, while dots are obtained in the Landau-Zener approximation. In all plots we set interaction strength $g=20$. The transition point $\lambda_0=6.365$. \label{Fig7}}
\end{figure}

Finally, let us also address an issue of the universality of our findings. This question is important since it is not excluded that the transition protocol proposed may be highly biased by the external potential chosen. Indeed, up to now, the external potential \eqref{potential} was chosen to have exactly the same central frequency for both components. In experimental realization, however, this constrain is of course not achieved rigorously. Especially, this is the case if one considers components of substantially different elements having different spectroscopic properties. Therefore, to settle if this small but rigid detail has any importance, in the following we consider a completely opposite extreme case. Namely, we focus on the parabolic external potential being exactly the same for both components ( for convenience expressed in terms of $A$-particle units)
\begin{align} \label{potential2} 
V_\sigma(x) = \left\{
\begin{array}{cl}
\frac{1}{2}\lambda m_A x^2, & |x|<L, \\
\infty, & |x|\geq L
\end{array}
\right.
\end{align}
In this scenario, all atomic constituents experience exactly the same, linearly position-dependent force, independently of their mass. The first sign that physics of both considered scenarios should be the same comes from the direct inspection of the many-body spectrum. As explicitly shown in Fig.~\ref{Fig7}a, the spectrum of the system with $N_A=1$ and $N_B=3$ particles resembles the spectrum of the system with the previous confinement (compare to the top plot in Fig.~\ref{Fig1}). Therefore, one can suspect that also the dynamical behavior of these two extreme cases will be very similar. Indeed, by repeating all the calculations in the same manner but for the current confinement, we obtained transition probabilities 
${\cal P}_0(\tau/2)$, ${\cal P}_1(\tau/2)$, ${\cal P}_0(\tau)$ presented in Fig.~\ref{Fig7}b (red and blue dots, respectively) and compare them with analytical predictions based on the Landau-Zener approximation. Exactly as previously, both approaches match almost ideally and they give additional premise that the protocol presented, when tailored appropriately, has a large dose of generality. We checked that similar conclusions are reached when the opposite imbalance case with $N_A=3$ and $N_B=1$ particles is considered. 

\section{Conclusions} \label{Sec6}
In this work, we have explored the dynamical properties of the interacting system of a few fermions driven through the critical transition by time-varying external potential. In contrast to previous studies \cite{2016PecakPRA,2020WlodzynskiPRA}, we focused on diabatic transition in which the system can be dynamically excited to other many-body states. By detailed investigation of temporal many-body spectra, we argued that a whole driving process can be well-described in significantly reduced Hilbert space spanned by only a few of the lowest many-body states. Particularly, in the case of an imbalanced system of four particles, only two many-body states are sufficient to give a comprehensive and adequate description of the system's behavior. In this case, a straightforward implementation of the Landau-Zener model, without any fitted parameters, gives a fully correct description of the many-body system's properties independently of the details of the external confinement. For a balanced system, the situation becomes less trivial. Nonetheless, characterization in terms of a few of the lowest states still gives complete knowledge on dynamical properties. All of this means that these particular systems of several and well-controlled particles may help in our understanding of systems containing a mesoscopic number of particles \cite{2012BlumeRPP,2016ZinnerEPJ,2019SowinskiRPP} and thus give some other perspective to many-body generalizations of the Landau-Zener problem \cite{2000WuPRA,2008AltlandPRL}.

We have argued that periodic driving of the system studied gives a route to coherent control on many-body excitations in systems of a small number of particles. By appropriate experimental tailoring of driving period and halting time, one can tune the final state of the system to a desired many-body eigenstate or their superpositions. Since the many-body eigenstates are typically not the states with extremal (maximal or minimal) quantum correlations, the method may be further exploited for the initial preparation of many-body states with assumed inter-component entanglement and then used for other purposes.

Finally, presented studies can be also utilized in an opposite direction, {\it i.e.}, to determine the system's parameters whose precise adjustment can be experimentally challenging, for example, the interaction strength $g$. Typically, for dynamical systems correctly modeled by the Landau-Zener transition, the energy gap in the transition point can be measured experimentally by measuring the transition probability \cite{2000WernsdorferJAPhys,2013UrdampilletaPRB}. In the case studied, the transition probability can be obtained from the experimentally accessible density profile since the ground state and the excited state have significantly different (opposite) separations of their components. Having the probabilities in hand, one can determine the interaction strength $g$, since there is a direct one-to-one relation between the interaction strength $g$ and the energy gap at the transition point. From this perspective, this additional possibility of determining the strength of interactions may have importance for quantum engineering. 

\acknowledgments
This work was supported by (Polish) National Science Centre Grant No. 2016/22/E/ST2/00555.

\bibliography{_BibTotal}

\end{document}